\documentclass[twocolumn,showpacs,preprintnumbers,amsmath,amssymb]{revtex4}

\usepackage{epsfig}
\usepackage{natbib}
\usepackage{amssymb}
\usepackage{latexsym} 
\usepackage{mathptm}

\def\beq{\begin{equation}}
\def\eeq{\end{equation}}
\def\beqn{\begin{eqnarray}}
\def\eeqn{\end{eqnarray}}

\begin{document}


\title{Deformed Special Relativity in Position Space}

\author{Sabine Hossenfelder}
  
\email{sabine@perimeterinstitute.ca}
\affiliation{Perimeter Institute, 31 Caroline St. N, Waterloo, Ontario, N2L 2Y5, Canada}
  
\begin{abstract}
We investigate how deformations of special relativity in momentum space can be extended
to position space in a consistent way, such that the dimensionless contraction between
wave-vector and coordinate-vector remains invariant. By using a parametrization in terms of
an energy dependent speed of light, and an energy dependent Planck's constant, we are
able to formulate simple requirements that completely determine the active transformations 
in position space. These deviate from the standard 
transformations for large velocities of the observed object. Some examples are discussed, 
and it is shown how the relativistic mass gain of a massive particle is affected. We 
finally study the construction of passive Lorentz-transformations.
\end{abstract}

\keywords{Deformed Special Relativity, Phenomenological Quantum Gravity}  
\pacs{11.10.Gh, 11.30.Cp, 12.90.+b}
 
\maketitle

\section{Introduction}
The Planck length is generally expected to act as a regulator in the ultraviolet, or as a 
fundamentally finite length respectively. Consequently, the corresponding Planck mass should be an
observer independent energy scale. The requirement that Lorentz-transformations in momentum
space leave this scale invariant leads to a class of deformations of special relativity ({\sc DSR}).
As one of the most general expectations from a theory of quantum gravity, these modified
Lorentz transformations have received much attention in the last years \cite{Amelino-Camelia:2000ge,Amelino-Camelia:2000zs,Jacobson:2001tu,Amelino-Camelia:2002vw,Sarkar:2002mg,Konopka:2002tt,Alfaro:2002ya,Amelino-Camelia:2002vy,Heyman:2003hs,Jacobson:2003bn,Magueijo:2001cr,Magueijo:2002am,Ahluwalia-Khalilova:2004dc,Smolin:2005cz,bloederdepp}. We will in the following refer to {\sc DSR} as the class 
deformations that has been examined in this context, and - maybe more accurately - has also been 
named 'Doubly Special Relativity', pointing towards the presence of two observer invariant scales.

Despite the fact that it is possible to use kinematic arguments to predict 
threshold corrections, a fully consistent quantum field theory with DSR is still not available.
Though there are notable attempts \cite{Hossenfelder:2003jz,Hossenfelder:2006cw,Konopka:2006fh,Magueijo:2006qd}, one of
the obstacles on the way is a formulation of {\sc DSR} in position space, which is essential for a 
meaningful interpretation of the action principle\footnote{One  can nevertheless
write down the action and derive the equations of motion, as well as Feynman rules in momentum
space, without actually knowing the transformations in position space. This however, is somewhat
unsatisfactory.}. The understanding of {\sc DSR} in position space is also
crucial for the derivation of conservation laws from space-time symmetries \cite{Hossenfelder:2006cw,Magueijo:2006qd,Agostini:2006nc}.

There are so far two approaches to the problem. The one requires a  non-commutative 
structure on the phase space \cite{Lukierski:1991pn,Majid:1994cy,Lukierski:1993wx,Kowalski-Glikman:2001gp,Bruno:2001mw,Kowalski-Glikman:2002we,Girelli:2005dc},
the other one results in an energy dependend metric \cite{Magueijo:2002xx,Kimberly:2003hp,Hinterleitner:2004ny,Amelino-Camelia:2005ne,Galan:2004st,Galan:2005ju,Hackett:2005mb,Galan:2006by,Ling:2006ba}. The latter formulation 
is quite intuitive if one keeps in mind that a particle causes space-time to be distorted through its energy, and the background therefore should strictly speaking 
be a function of the particle's energy. This is nothing but a consequence of the backreaction problem in general
relativity \cite{Hossenfelder:2006cw}. However, as shown in \cite{Kimberly:2003hp}, this notion of an
energy dependent metric results in a confusing definition of the relative velocity, and an explicit 
expression for the transformations remains to be given. In particular, one question that one would
like to clarify is how a speed of light that can approach infinity in some models of {\sc DSR} can not
single out a preferred notion of equal time slices. Or, to put it differently, how can all observers
with relative motion agree on the speed of light being infinite? Another question arising from
this is whether a massive particle still experiences an upper bound on its speed if the speed of 
light can approach infinity.

As we will show in the following, the physical definition of the relative velocity of reference
frames can be used to fix the ambiguity in the parametrization of Lorentz transformations. By
keeping track of the dimensionality of quantities, we show that conjugated quantities can transform
appropriately without the need to introduce an energy dependent metric, or a non-commutative
geometry.  

As pointed out in this earlier work, {\sc DSR} can either be seen 
as a theory that effectively describes particle interactions in space-time regions of 
non-negligible curvature -- in which case only the virtual particles are
subject to {\sc DSR} -- or the single free particle's properties are also described by {\sc DSR}. 
In the latter interpretation, the momentum of free particles becomes a non-additive
quantity, which leads to conceptual problems in the formulation of a field theory, such as
the proper definition of conserved quantities in interactions, and the transformation of multi-particle
states (also known as the soccer-ball problem). A general goal of our investigation should be to decide which of the two interpretations that were discussed in \cite{Hossenfelder:2006cw} can be pushed forward to eventually result in a cleanly 
defined quantum field theory. The aim of this paper is to focus on the observer-independent description
in position space. 

This paper is organized as follows. In the next section we recall the formalism of {\sc DSR} in momentum
space, and clarify the notation that we will use. In the third section, we will formulate some
requirements to extend the symmetry to position space. We start with deriving the active transformations and 
examine some examples. We then turn towards the passive transformations, followed by a discussion of the
results.

Exceptionally, $\hbar$ and $c$ are not equal to one, and will be carried through the paper. 
$m_{\rm p}$ is some mass scale, to be identified with the Planck mass. 

\section{Deformation in Momentum Space}

We will use the notation introduced in \cite{Hossenfelder:2003jz,Hossenfelder:2006cw,Hossenfelder:2005ed}. 
The quantity ${\bf p} = (E,p)$ transforms as a standard Lorentz-vector, and is distinguished from 
${\bf k} = (\omega,k)$, which obeys the modified transformation that is non-linear in 
${\bf k}$. The former quantity {\bf p} can always be introduced, the important step is eventually its 
physical interpretation. As investigated in \cite{Hossenfelder:2006cw}, there are two conceptually different ways
to attach physical meaning to these quantities. In the approach of \cite{Hossenfelder:2003jz}, ${\bf p}$ 
is the physical momentum four-vector, to be distinguished from the wave-vector ${\bf k}$. Whereas in 
the more common approach ${\bf k}$ plays the role of both, momentum and wave-vector, 
and ${\bf p}$ is a pseudo-variable, useful for calculations, but void of physical content. The following
formalism can be used for both interpretations.

A common notation for the relation between {\bf p} and {\bf k} is
\beqn
\omega = E f(E) \quad,\quad k = p g(E)\quad. \label{trafo1}
\eeqn
This does cover the most common DSR realizations, but a general relation can be of the form
\beqn
{\bf k} = F({\bf p}) = \left( E f({\bf p}), p ~g({\bf p})\right)\quad,
\eeqn
with the inverse ${\bf p} = F^{-1}({\bf k})$. As examined in \cite{Hossenfelder:2005ed} 
these theories can, but need not necessarily have an energy dependend speed of light. I.e., since
the relation is invertible one sees immediately that a function of the form 
$(\omega,k) = (Eg(E), p g(p))$, which has been used e.g. in \cite{Hossenfelder:2003jz}, does 
not modify the light-cone, a choice that is unfortunately not covered by the parametrization (\ref{trafo1}).

An obvious requirement is that the function $F$ reduce to multiplication with $\hbar$ in the limit
of energies being small with respect to the Planck scale. In order to implement a maximum energy
scale, either one or all components of {\bf k} should be bounded by $m_{\rm p}$. Since we eventually are interested 
in constructing a quantum field theory which respects the deformed Lorentz-symmetry as well 
as {\sc CPT}-symmetry, we should further demand that $F$ be an odd function $F(-{\bf p}) = - F({\bf p})$.
 
For the following it will be useful to recast the notation in two functions that play
the role of an energy dependend speed of light, and an energy dependent Planck's constant 
\beqn
\tilde c({\bf p}) = \frac{\omega}{k} = c \frac{f({\bf p})}{g({\bf p})} \quad, 
\quad \tilde \hbar({\bf p}) = \frac{1}{f({\bf p})} \quad, 
\eeqn
such that we have
\beqn
E = \tilde \hbar \omega \quad, \quad  c p = \tilde \hbar \tilde c k \quad. \label{trafo2}
\eeqn
It is well known that under quantization, these modifications lead to a generalized uncertainty principle
\cite{Hossenfelder:2005ed,Kempf:1994su,ml1,ml2,ml3}. 
One should keep in mind that $\tilde c$ is just a general function of the energy of the particle under 
consideration, which has the property that for a photon with $E_\gamma = c p_\gamma$, it coincides with
the photon's energy dependend speed. The constant $c$ is a parameter which agrees with the speed of
light in the low energy limit and is thus our familiar constant $c$. For a massive particle, 
the relations Eqs. (\ref{trafo2}) however, will not be a function of some photon's energy, but of 
the particle's energy itself. In particular, one notices that even in the restframe, $p=0$, of a 
massive particle $\tilde c$ is not identical to $c$. Instead, c is
multiplied by a function of $m^2$, where first  deviations\footnote{Provided the DSR respects the
above mentioned symmetry under ${\rm p}\to -{\rm p}$.} for $\tilde c/c -1$ will be suppressed with
a power of $m^2/m_{\rm p}^2$.

It's been known for some while how to achieve a transformation that maps $\bf k \to {\bf k}'$ 
and respects the invariance of the modified dispersion relation
\beqn
(\tilde \hbar \tilde c k)^2 - (\tilde \hbar \omega )^2 = (\tilde \hbar' \tilde c' k')^2 - (\tilde \hbar' \omega' )^2 \quad, 
\eeqn
where we use the notation $\tilde \hbar' = \tilde \hbar({\bf p}')$, $\tilde c' = \tilde c ({\bf p}')$. Or, since 
$F$ is invertible this can also be read as   
$\tilde \hbar' = \tilde \hbar({\bf k}')$, $\tilde c' = \tilde c ({\bf k}')$. In fact, these transformations
are straight forward to derive for a given $\tilde c, \tilde \hbar$. One just keeps in mind that the relation 
for $(E,p)$ is the standard relation
\beqn
(c p)^2 - E^2 = (c p')^2 - E'^2 \quad,
\eeqn
from which one finds the standard Lorentz transformation in momentum space:
\beqn
E' = \gamma_w (E -  w p) \quad, \quad p' = \gamma_w (p-\frac{w}{c^2} E) \quad.
\label{lotrap}
\eeqn
where $1/\gamma_w = \sqrt{1 - w^2/c^2}$.
Note that so far, the $w$ that appears in this equations is nothing but a parameter that labels the transformation.
It doesn't yet have a physical meaning. The index on the $\gamma_w$ indicates for later convenience 
that $\gamma_w$ strictly speaking also is a function of the parameter $w$ that labels the transformation. 
We will denote these transformations with ${\bf p}' =  L(w,{\bf p})$. 

Now one gets the modified  Lorentz-transformations acting on {\bf k} by 
\beqn 
{\bf k}' = F({\bf p}') = F( L(w,{\bf p})) = F(L(w, F^{-1}({\bf k}))) \quad,
\eeqn
which yields
\beqn 
\omega' &=& \frac{E'}{\tilde \hbar'} = \gamma_w (\omega - w \frac{\tilde c}{c} k) \frac{\tilde \hbar}{\tilde \hbar'} \nonumber \\
k' &=& \frac{c p'}{\tilde c \tilde \hbar}  = \gamma_w (k - \frac{v}{c \tilde c} \omega) 
\frac{\tilde c}{\tilde c'}\frac{\tilde \hbar}{\tilde \hbar'} \quad. \label{trafok}
\eeqn
 The transformations Eqs. (\ref{trafok}) will be non-linear 
in $(\omega,k)$ since $\tilde c$ and $\tilde \hbar$ are functions of these quantities as well. By construction,
the transformations respect implemented upper bounds on one or all components of {\bf k}. For special choices of
$\tilde c$ and $\tilde \hbar$ one finds the {\sc DSR} transformations used in the literature. We will denote
these transformations as ${\bf k}' = \widetilde{L}(w,{\bf k})$. 

One known problem with this approach is that bound systems of elementary particles can
very well exceed the Planck mass, and the transformations therefore can not apply for them. The reason
for this mismatch, also known as the soccer-ball problem, is the non-linearity of the 
transformations, which should be suppressed when the number of constituents grows.
Unfortunately, even if a classical argument was available, it could not easily be transferred to
quantum systems, in which the total number of constituents is a very ill defined notion due to virtual
particle content. Though large progress has been made regarding the solution of this
problem \cite{Magueijo:2006qd,Hinterleitner:2004ny,Judes:2002bw,Girelli:2004ue,Girelli:2006ez,Hossenfelder:2007fy}, the issue is still not completely settled
and open questions remain \cite{Ahluwalia-Khalilova:2004dc,Liberati:2004ju,Toller:2003tz}.

For such multi-particle systems it is then not a-priori clear how to generalize the here used approach
with constants being modified to energy dependent functions, as it is not clear which energy these functions
should depend on.  Throughout this paper however, we will deal with single particles for which the energy 
on which these functions depend is just the particle's energy. 

It should also be noted that the soccer-ball problem is not present in the interpretation
of {\sc DSR} given in \cite{Hossenfelder:2006cw}. In this case, modifications do only arise if the total energy
of a system, or an interaction taking place, causes a strong enough background curvature to make quantum
gravitational effects non-negligible. This is in general not the case for typical systems bound through the
standard model interactions. In the following sections, we will examine the consistency of 
the more common interpretation, in which the free particle also is subject to the {\sc DSR} formalism.

\section{Deformation in Position Space}

If one wants to construct a field theory that respects a deformed Lorentz-symmetry, it is essential 
for the Lagrangian formulation to have the corresponding transformation in position 
space. It is also necessary to understand the properties in position space in order to obtain 
conserved Noether charges that can be derived from the symmetry principles. One might argue 
that knowledge of the position space formulation  is not necessary to arrive at the 
threshold corrections which arise in some formulations of {\sc DSR}. But a position space 
description definitely is necessary if one wants to investigate whether a possible energy 
dependence of the time of flight is present, and measurable, e.g. for high energetic photons 
from $\gamma$-ray bursts. For a recent evaluation of the detectability see e.g. \cite{Scargle:2006kr}. 

The approach pursued in \cite{Kimberly:2003hp} is way leading, but it still remains
the question how the transformations explicitly are constructed. The resulting final 
transformation between two reference frame with a relative velocity $v$ should be 
a function only of this parameter, as there is no other parameter in the game.
 
For completeness let us first recall the derivation of the usual Lorentz transformations in position space
\beqn
\Delta t' &=& \gamma (-\frac{v}{c^2} \Delta x + \Delta t) \quad, \quad
\Delta x' = \gamma (\Delta x - v \Delta t) \label{lotrax} \quad,
\eeqn
which one derives most easily as those transformation that leave the line element in Minkowski space invariant, 
i.e. the Lorentz transformations are just SO$(3,1)$, obviously. If one does so, one is left with one 
free parameter which can be fixed by requiring 
that the origin of the unprimed space moves with velocity $-v$, i.e.
\beqn
\frac{\Delta x'}{\Delta t'} (\Delta x = 0) = -v \quad. \label{relvel}
\eeqn
Now let us construct the modified transformations in position space that close with the above defined transformations on
{\bf p}, or {\bf k} respectively (since we have shown that the one implies the other if $\tilde c$ 
and $\tilde \hbar$ are given). 


Let us start with considering active transformations. The expected modification of the standard requirement is that for an
active boost, there is no such constant as $c$. Instead $\tilde c$ itself also transforms and 
turns into $\tilde c'$. In addition, one has first to make sure that the contraction of
the examined quantity with {\bf p} results in a dimensionless scalar. That is, the quantities 
need to have dimension of an inverse energy \footnote{Note also that 
$\omega^2 - (\tilde c k)^2$ is {\sl not} invariant, and therefore its canonically conjugated 
quantities should not be considered here.}. 
The requirement of invariance in space-time is now
\beqn
\left(\frac{\Delta x}{\tilde \hbar \tilde c}\right)^2 - \left(\frac{\Delta t}{\tilde \hbar}\right)^2 = 
\left(\frac{\Delta x'}{\tilde \hbar' \tilde c'}\right)^2 - \left(\frac{\Delta t'}{\tilde \hbar'}\right)^2 \quad. 
\label{modreq}
\eeqn
Since we are 
considering active transformations, we are describing a particle moving in a restframe with a relative motion $v$, 
and are asking how the particle's properties transform with the parameter $v$, which is an observable. We 
assume that we are considering the particle's motion on macroscopic scales considerably above the 
Planck scale, such that its location (but not necessarily its energy or velocity) has the standard, 
classical, properties. 
The particle's position should thus be interpreted as an expectation value. 
 
We are lead to the transformations   
\beqn
\Delta t' &=& \tilde \gamma_v (-\frac{v}{\tilde c \tilde c'} \Delta x + \Delta t) \frac{\tilde \hbar'}{\tilde \hbar} \quad,\\
\Delta x' &=& \tilde \gamma_v (\Delta x - v \frac{\tilde c}{\tilde c'} \Delta t) \frac{\tilde \hbar'}{\tilde \hbar} 
\frac{\tilde c'}{\tilde c} \label{lotraxnew} \quad,
\eeqn
where $1/\tilde \gamma_v = \sqrt{1- v^2/\tilde c'^2}$, 
which arises since we were now able to fix the parameter in the transformations by using the consistency
requirement Eq.(\ref{relvel}). 
In these equations, the quantity $\tilde c'$ carries the information
about the particle's properties\footnote{To set the Eqs.(\ref{lotraxnew}) in relation to the transformations in \cite{Kimberly:2003hp} Eq.(15), 
note that in the latter transformations, the parameter $v$ is not the relative velocity 
(as mentioned in the paper later on). In particular one has
$ \Delta x' / \Delta t' = v (g(E')g(E))/(f(E)f(E'))$.
With a suitable redefinition of the parameter $v$, the transformations are identical to those derived above.}. 
The transformations relate a particle with restmass $m$ in a restframe 
to the same particle with velocity $v$ relative to the restframe. 

However, these transformations
aren't yet fully satisfactory.  First, we are left with
fixing the parameter $w$ in the momentum space transformations Eqs.(\ref{trafok}). For this we have
to relate the position space with the dual space, and so we examine the invariance of the 
contraction  
$- \Delta t \omega + \Delta x k =   - \Delta t' \omega' + \Delta x' k'$.
From this one can easily identify the parameter $w$ in Eqs. (\ref{trafok}) to be $w = v c/\tilde c'$. So, 
the transformations in momentum space finally read:
\beqn 
\omega' &=&  \tilde \gamma_v (\omega - v \frac{\tilde c}{\tilde c'} k) 
\frac{\tilde \hbar}{\tilde \hbar'} \quad, \nonumber \\
k' &=&  \tilde \gamma_v (k - \frac{v}{\tilde c \tilde c'} \omega) 
\frac{\tilde c}{\tilde c'}\frac{\tilde \hbar}{\tilde \hbar'}  \label{trafok2}\\
E' &=& \tilde \gamma_v (E -  v \frac{c}{\tilde c'} p) \quad , \nonumber \\ 
p' &=& \tilde \gamma_v(p-\frac{v}{c \tilde c'} E) \quad.
\label{lotrap2}
\eeqn
For massless particles the transformations are now completely determined if one knows the
energy of the particle, since the very definition of $\tilde c$ provides us with a direct, and invertible,
relation between the massless particle's energy and its speed. But for a massive particle, 
we still need to know $\tilde c'$ in the boosted frame, which requires knowledge about the transformation
of the energy as a function of the relative velocity. 

The question that we are facing now is, given that we define $\tilde c$ in one restframe,
then how does this look like in another reference frame that moves relative
to the first with velocity $v$? That is, what we need is $\tilde c' = \tilde c(m,v)$, which removes
the energy dependence of the transformation. 

This missing 
relation is already implied by the {\sc DSR} transformations in the form
\beqn
\tilde c'&=& \tilde c ( {\bf p}') = \tilde c (L(v,{\bf p})   \quad,\\ 
\tilde \hbar' &=& \tilde \hbar ({\bf p}') = \tilde \hbar (L(v, {\bf p} )) \quad, \label{impl2}
\eeqn
where {\bf p} can be chosen as $(m,0)$, and the functions $\tilde \hbar$ and $\tilde c$ are the input of the
theory.  The complication which arises is 
that $L(v,{\bf p})$ is again a function of $\tilde c'$ and $\tilde \hbar'$, so 
the Eqs. (\ref{impl2}) are two implicit equations for the desired two quantities, which will in general be
hard to solve. In principle however, the requirement that $\tilde c'$ has to fulfill these equations gives the desired relation between
$\tilde c'$, $m$ and $v$, which one can eventually insert 
in Eqs. (\ref{lotraxnew}) to obtain the Lorentz-transformations
in position space. 


If one closes the transformations with these requirements, the transformations in position space
are no longer energy dependend for a massive particle, since the relativistic energy can be expressed
through the particle's rest mass and its relative velocity. For a massless particle, the transformation depends on the 
particle's energy. By construction (\ref{modreq}), the energy dependence of the massless particle's speed is
respected by the transformations.

It is also straight forward to derive the addition law for velocities. With the notation 
$\tilde c'' = \tilde c(w), \tilde \hbar'' = \tilde \hbar(w)$ one finds
\beqn
u = \frac{\frac{v \tilde c''}{\tilde c'}+w}{1+\frac{vw}{\tilde c\tilde c''}} \quad.
\eeqn
Since $\tilde c$ is a function of the velocity, this is not the standard addition law\footnote{Which thus explains
the difference between the approach presented here, and the one discussed in \cite{Kosinski:2002gu}, where the
Einstein addition law was used to define the velocity.}. One verifies easily that a particle moving with the
speed of light ($v=\tilde c'$, or $w=\tilde c''$) does so still after applying an additional boost 
(it follows $u=\tilde c''$). The new transformations still form a representation of the Lorentz-group, since they are
isomorphic to the standard tranformation under applying the map provided by the relation between the physical velocity 
and the parameter labeling the standard transformations.

But maybe most importantly, one sees that it is possible to formulate {\sc DSR} in position space without the
need to use non-commutative geometries, or an energy dependent metric. It is also interesting  that a 
modification of the Lorentz-transformation occurs both for ${\bf p}$ as well as for ${\bf k}$ if we demand $x,t$ to
be usual coordinates, from which we can sensibly define a relative motion. The underlying reason for this is 
that (from dimensional
arguments) it is {\bf k} that generates the translations in space-time\footnote{As developed earlier \cite{Hossenfelder:2003jz}, under quantization
it is therefore {\bf k} that becomes the partial derivative operator. Consequently, after quantization, 
one obtains higher order operators for ${\bf p}$ which
generates the wave equation, and give the propagator of a quantum field theory.}. Since it has been argued that
predictions of {\sc DSR} can always be made to vanish by a redefinition of parameters \cite{Jafari:2006rr}, 
it is worthwhile to stress that this result shows that such a redefinition does not remove the deformation
of the transformation, since an interpretation of the parameters relies on a proper definition of the
relative velocity. 
If one starts with a modified dispersion
relation for the wave-vector that contracts with space-time coordinates, a redefinition to 
pseudo-variables will not suffice to declare the theory void of context. As one can see from the argument 
above, in this case the pseudo-variables will obey modified transformation behavior as well, since the 
parameter of a relative velocity appearing in these transformations is still connected to the modified quantity which is
contracted with the coordinate vector.


Let us briefly summarize the assumptions that we made to arrive at this result. First, though we did not
explicitly state it, we assumed that the {\sc DSR} has the same form for all kinds of particles. 
We have further assumed that the scalar product in position space Eq.(\ref{modreq}) takes into 
account that the speed of light, and Planck's constant, also transform under a change of the relative velocity. 
The importance of Planck's constant arises from the need to connect conjugated quantities. 
We were left with
one free parameter that we could express through the relative velocity. A consistency requirement on the contraction
over position and the dual quantities eventually allows us to express the energy dependence of $\tilde c$ and
$\tilde \hbar$ as a velocity dependence.  If one likes, 
one can pull out the additional factors in Eq.(\ref{modreq}), and define them to be content of a 
modified metric, which then coincides with the approach in \cite{Kimberly:2003hp}. The components of
the metric can alternatively be expressed as functions of the particle's energy, or its velocity.
One should note that the soccer-ball problem stays with us when we go from momentum to 
position space.  
 

So far, we have examined active transformations that relate particles with different
velocities. We will now turn towards the question: is it possible to construct passive 'deformed' 
Lorentz transformations acting on our space-time coordinates, such that an energy dependend speed 
of light $\tilde c(E)$ is allowed to remain observer independent?

With a passive transformation we mean a change of coordinates, as opposed to an active transformation which changes
the velocity of an observed object, but not the coordinates it is described in. The latter possibility is
what we have examined in the previous section. The transformations that we are looking for now 
are deformed in the sense that they respect the postulated 
{\sc DSR} transformations in momentum space, which give a prescription on 
how $E$ relates to $E'$. With observer independent we mean that all observers agree on the 
relation $\tilde c(E)$. That is, if one observer
sees a particle with $E$ and $\tilde c(E)$, then the boosted observer who measures the particle having 
energy $E'$  measures its speed to be $\tilde c(E')$. 

We will also require observer independence to include 
independence of the space-time location. This is a requirement that can be altered if one wants to 
use a background that explicitly breaks the maximal symmetry of Minkowski space. We will come back
to this in the discussion in section \ref{dis}.

We are looking for a passive transformation $\Lambda_{\rm P}$ that takes the 
coordinates of restframe A with $(t,x)$ to that of restframe B with $(\tau,\chi)$. Independence of the physics
on the space-time location means that the transformation has to be linear. If it wasn't, it would depend on
$t$ and/or $x$ and break homogeneity of time, and/or homogeneity and isotropy of space. For convenience, we will
then in the following talk about space and time distances to an arbitrary point $t_0,x_0$ in restframe A 
that corresponds to $\tau_0,\chi_0$ in restframe B, and use the distances of coordinates to this point
$\Delta t, \Delta x$, and $\Delta \tau, \Delta \chi$. Thus, the transformation can be written in matrix notation as
\beqn
\Lambda_{\rm P} =  \left( 
\begin{array}{cc}
\Lambda^\tau_{\;\;t} &  \Lambda^\tau_{\;\;x}\\  
\Lambda^\chi_{\;\;t}& \Lambda^{\chi}_{\;\;x} 
\end{array} \right) \quad,\quad 
\Lambda_{\rm P}  \left( 
\begin{array}{cc}
\Delta t \\  
 \Delta x
\end{array} \right) =  \left( \begin{array}{cc}
\Delta \tau \\  
 \Delta \chi
\end{array} \right)\quad. 
\eeqn
Again, we keep in mind that space-time coordinates, or
distances, might loose a distinct meaning close by the Planck length. Therefore let us stress again that we are
considering macroscopic effects, as for example the time of flight analysis for $\gamma$-ray bursts 
within {\sc DSR} where $\Delta x 
\sim$Gpc, and particles in the space-time with energies of $E\leq m_{\rm p}$. For such scales, coordinates 
and distances should exist and behave as usual. For a more detailed argument, see the discussion in
section \ref{dis}.

We are examining passive transformations which are parameterized by the relative velocity of reference 
frames $v\neq 0$, the parameter being given by $\Delta x/\Delta t(\Delta \chi=0) := - v$.
From this it follows that
\beqn
\Lambda^\chi_{\;\;t} = v \Lambda^\chi_{\;\;x} \label{3} \quad. 
\eeqn
We further require $\Lambda_{\rm P}$ to be invertible, as not to result in a lower dimensional target space. 
Now to the photons. In restframe A two photons with energies $E_1$ and $E_2 \neq E_1$ have to obey
\beqn
\frac{\Delta x}{\Delta t} = \tilde c(E_1) \quad, \quad \frac{\Delta x}{\Delta t} = \tilde c(E_2) \quad, \label{4}
\eeqn
where $\Delta x$ could either be positively or negatively valued, and we have chosen both $\Delta x$ to be the same. 
I.e. the photons are received after they traveled the same distance in restframe A, but with a possibly 
different time span.
In restframe B we have then
\beqn
\frac{\Delta \chi_1}{\Delta \tau_1} = \tilde c(E'_1) \quad, \quad \frac{\Delta \chi_2}{\Delta \tau_2} = \tilde c(E_2') \quad, \label{5}
\eeqn
where the relation between $E_{1,2}$ and $E'_{1,2}$ is given by the DSR transformation and in principle known. We 
will not need it explicitly, but we notice that $\tilde c$ should be a smooth and monotonically increasing 
function, as to be invertible and thus from $E_1 \neq E_2$ it follows $c(E_1) \neq c(E_2)$. 
To arrive at these equations for different photons, it is not necessary to actually have a system with both photons
present. 

From Eqs.(\ref{4}) and (\ref{5}) one gets
\beqn
\frac{\Lambda^\tau_{\;\;t}}{\tilde c(E_1)} + \Lambda^\tau_{\;\;x} - \frac{1}{\tilde c(E_1')}
\left( \frac{\Lambda^\chi_{\;\;t}}{\tilde c(E_1)} + \Lambda^\chi_{\;\;x} \right) = 0 \quad, \label{8}
\eeqn
that is one constraint on the entries of $\Lambda_{\rm p}$.

We could repeat this procedure for $n$ different photons of different energies, which will
inevitably bring us into trouble with the construction of a linear transformation. Let us do a 
quick counting of variables to examine the issue of determining the matrix that we
are looking for. $\Lambda_{\rm P}$ has four entries. Each equation for one photon gives two equations. 
For $n$ photons with energies $E_1,E_2,...E_n$ one had $2n$ equations. But each photon also adds one
unknown $\Delta \chi_n$. Taken together this means measurement of $n$ photons of different energies yield
$2n$ equations for $4+n$ unknown variables.

Thus, the transformations were already fully determined if one would considers $4$ photons 
of different energy,  and over determined for all additional photons of which we potentially
have infinitely many. 
These equations can be fulfilled if and only if infinitely many of the equations of the form
(\ref{8}) for all possible energies are linearly dependend. Taken together with the requirement 
that $\tilde c$ be a smooth function this means $\tilde c$ has to be constant.

\subsection{Discussion}
\label{dis}

The results of the previous section indicate an inherent conflict between energy dependence of
the speed of light, and observer independence. As pointed out before, it is of course possible 
to explicitly break observer independence to allow an energy dependent speed of light. Caused by a preferred
background frame (e.g. given by the CMB), one can postulate $\tilde c(E)$ 
to hold only in one frame (e.g. the one in rest with the CMB). Since it doesn't really make
sense to say in any other frame photons don't travel with the speed of light, let us instead 
formulate it as: in every other frame the relation $\tilde c(E)$ is modified. 
These conclusions seem to favor versions of {\sc DSR} that have an energy dependend Planck's constant, but
not also an energy dependent speed of light.
 
One should note that the above derivation rests on a fairly general argument. If one has a 
linear transformation acting on a space, one can't take arbitrarily many independent 
directions and require them to transform according to a given function (in this case $\tilde c$). 
In case $c$ is constant, all of the photons trajectories are linearly dependent, and thus 
the system is fully determined, but not over determined. 

It is in the above made assumption that it matters whether we consider 
an active or a passive boost. For the passive boost, we are looking for a transformation 
between the two coordinate systems A and B. For an active boost on the contrary, we were considering 
$n$ different particles, whose transformation properties can in principle be different. In particular, as
we have seen it turned out that the transformations
depend on the particle's energy. 

Now, for the passive boosts in principle a particle with energy $E$ would cause
a distortion of the geometry and thereby influence the measurement of distances. Strictly speaking,
the transformation $\Lambda_{\rm P}$ therefore should also be a function of the particle's
energy. For the same reason, the transformation should strictly speaking not be linear,
since the very presence of the observed particle breaks homogeneity and causes a position 
dependence. One might call this an observation dependence of the coordinate transformation.
However, since we are considering macroscopic distances and particles with energies
below the Planck scale, such observation dependent effects are negligible, as can easily be estimated. 

Consider a particle of energy  $E \leq m_{\rm p}c^2$ somewhere in the coordinate 
system, say at position $x_E$.  We might not know what the particle's gravitational 
field looks like at a Planck scale distance, but we know it obeys the laws of General 
Relativity for distances far above the Planck scale. In
particular, the potential of the particle vanishes for $|x-x_E|\gg l_{\rm p}$ 
like $(E l_{\rm p})/(m_{\rm p} c^2 |x-x_E|)$. Let us place the particle in the middle of the distance $\Delta x$,
and cut out the quantum gravitational region of a size $\sim l_{\rm p}$. Inside this region,
the coordinate distance gets distorted to an unknown distance that we will
denote $d_{\rm QG}$, and which should not be of macroscopic size. 

Outside this quantum gravitational region, we integrate over the potential and find the distortion
\beqn
\Delta x &\to& d_{\rm QG} + 2  \int_{l_{\rm p}/2}^{\Delta x/2} dx \left( 1 - \frac{E l_{\rm p}}{m_{\rm p}c ^2 x} \right) \nonumber\\
&=& \Delta x \left( 1 + \frac{d_{\rm QG}- l_{\rm p}}{\Delta x} + 2 \frac{E l_{\rm p}}{m_{\rm p} c^2 
\Delta x} \ln(\Delta x/l_{\rm p}) \right)~, \label{estimate}
\eeqn  
where the term in the brackets gives the order of non-linearity and energy dependence 
that we can expect for $\Lambda_{\rm P}$.
Here, the logarithmic contribution is due to the fact that strictly speaking the potential of the
particle never vanishes exactly. 

We thus have two contributions. The one is an absolute additional unknown term of the
order of the Planck length that, no matter what its exact value, should eventually become
negligible for sufficiently large distances. This is a consequence of the fact that we have only
one particle, and its influence does not scale linearly with the distance. We further have a contribution that drops  
with $\ln(\Delta x)/\Delta x$, and is completely negligible relative to the leading order effect. I.e. 
even for a Gpc, the logarithm of $\Delta x/l_{\rm p}$ is only $\sim$ 60. As one expects, the correction
terms become non-negligible for very small distances $\Delta x \sim l_{\rm p}$.

This has to be contrasted with the modification of the passive boost that would be required to
enable observer independence of the energy dependent speed of light. If we allow a transformation
from one coordinate system to the other to depend on the observed particles energies, 
then we can fulfill the requirements formulated above in assumption three even though the speed of light
is energy dependent. 
The transformation that we would then inevitably be lead to were the same as the active boosts derived in the previous
section. That is, {\sc DSR} with an energy dependent speed of light would require a particle of mass $E\leq m_{\rm p}c^2$ to distort arbitrarily large
distances $\Delta x$ by a factor that does not depend on the distance, but only on the energy of the particle 
\beqn
\Delta x \to \Delta x \left(1 + {\mathcal O}\left( \frac{E}{m_{\rm p}} \right) \right) \quad, \label{dsrmod}
\eeqn
which is in strongest disagreement with our estimate Eq. (\ref{estimate}). In fact, one can reverse
the steps that lead to our estimate above with the modification Eq.(\ref{dsrmod}). Using an ansatz of
an arbitrary potential that should lead to Eq.(\ref{dsrmod}), one finds that
the particle's influence on the background does not drop with the distance to the particle. This
is a very unusual modification of the gravitational field which  should be 
motivated carefully.

\section{Conclusion}

We have examined an extension of deformations of special relativity from momentum to position space, and
have shown that a class of active boosts can be constructed that respects the momentum space symmetry. For 
this, it was not necessary to introduce non-commutative coordinates, or an energy dependent metric, 
though the approach presented here can be formulated in terms of the latter.  We 
have examined modifications of the relativistic mass gain of a massive particle in a theory 
with deformed Lorentz-symmetry, and shown that the particle's energy does not diverge for any finite 
speed. We have further examined how passive transformations can be constructed, and were
lead to the conclusion that this attempt is in conflict with an energy dependent speed of light.

\section*{Acknowledgements}

I thank Florian Girelli, Ulrich Harbach, Franz Hinterleitner, Stefan Hofmann, Benjamin Koch, Tomasz Konopka, and Lee Smolin for helpful discussions and criticism. Research at Perimeter Institute for Theoretical Physics is supported in
part by the Government of Canada through NSERC and by the Province of
Ontario through MRI.


\begin{thebibliography}{99}


\bibitem{Amelino-Camelia:2000ge} 
G.~Amelino-Camelia, 
Phys.\ Lett.\ B {\bf 510}, 255 (2001) [arXiv:hep-th/0012238]. 

\bibitem{Amelino-Camelia:2000zs} 
G.~Amelino-Camelia and T.~Piran, 
Phys.\ Rev.\ D {\bf 64}, 036005 (2001) [arXiv:astro-ph/0008107]. 

\bibitem{Jacobson:2001tu} 
T.~Jacobson, S.~Liberati and D.~Mattingly, 
Phys.\ Rev.\ D {\bf 66}, 081302 (2002) [arXiv:hep-ph/0112207]. 
 

 

\bibitem{Amelino-Camelia:2002vw} 
G.~Amelino-Camelia, 
Mod.\ Phys.\ Lett.\ A {\bf 17}, 899 (2002) [arXiv:gr-qc/0204051]. 

\bibitem{Sarkar:2002mg} 
S.~Sarkar, 
Mod.\ Phys.\ Lett.\ A {\bf 17}, 1025 (2002) [arXiv:gr-qc/0204092]. 
 
\bibitem{Konopka:2002tt} 
T.~J.~Konopka and S.~A.~Major, 
New J.\ Phys.\ {\bf 4}, 57 (2002) [arXiv:hep-ph/0201184]. 

\bibitem{Alfaro:2002ya} 
J.~Alfaro and G.~Palma, 
Phys.\ Rev.\ D {\bf 67}, 083003 (2003) [arXiv:hep-th/0208193]. 

\bibitem{Amelino-Camelia:2002vy} 
G.~Amelino-Camelia, 
Int.\ J.\ Mod.\ Phys.\ D {\bf 11}, 1643 (2002) [arXiv:gr-qc/0210063]. 

\bibitem{Heyman:2003hs} 
D.~Heyman, F.~Hinteleitner and S.~Major, 
Phys.\ Rev.\ D {\bf 69}, 105016 (2004) [arXiv:gr-qc/0312089]. 
 

\bibitem{Jacobson:2003bn} 
T.~A.~Jacobson, S.~Liberati, D.~Mattingly and F.~W.~Stecker, 
Phys.\ Rev.\ Lett.\ {\bf 93}, 021101 (2004) [arXiv:astro-ph/0309681]. 

\bibitem{Magueijo:2001cr}
  J.~Magueijo and L.~Smolin,
  Phys.\ Rev.\ Lett.\  {\bf 88}, 190403 (2002)
  [arXiv:hep-th/0112090].

\bibitem{Magueijo:2002am}
  J.~Magueijo and L.~Smolin,
  Phys.\ Rev.\ D {\bf 67}, 044017 (2003)
  [arXiv:gr-qc/0207085].

\bibitem{Ahluwalia-Khalilova:2004dc} 
D.~V.~Ahluwalia-Khalilova, 
Int.\ J.\ Mod.\ Phys.\ D {\bf 13}, 335 (2004) [arXiv:gr-qc/0402023]. 
 
\bibitem{Smolin:2005cz} 
L.~Smolin, 
[arXiv:hep-th/0501091]. 


\bibitem{bloederdepp}
J.~R.~Ellis, N.~E.~Mavromatos, D.~V.~Nanopoulos, A.~S.~Sakharov and
E.~K.~G.~Sarkisyan,
Astropart.\ Phys.\ {\bf 25} (2006) 402
[arXiv:astro-ph/0510172].

  
\bibitem{Hossenfelder:2003jz}
  S.~Hossenfelder {\sl et al},
  Phys.\ Lett.\ B {\bf 575}, 85 (2003)
  [arXiv:hep-th/0305262].
\bibitem{Hossenfelder:2006cw} 
S.~Hossenfelder, 
Phys.\ Rev.\ D {\bf 73}, 105013 (2006) [arXiv:hep-th/0603032]. 

 

\bibitem{Konopka:2006fh}
  T.~Konopka,
  arXiv:hep-th/0601030.

\bibitem{Magueijo:2006qd} 
J.~Magueijo, 
Phys.\ Rev.\ D {\bf 73}, 124020 (2006) [arXiv:gr-qc/0603073]. 

\bibitem{Agostini:2006nc}
  A.~Agostini, G.~Amelino-Camelia, M.~Arzano, A.~Marciano and R.~A.~Tacchi,
  arXiv:hep-th/0607221.

 
\bibitem{Lukierski:1991pn} 
J.~Lukierski, H.~Ruegg, A.~Nowicki and V.~N.~Tolstoi, 

\bibitem{Majid:1994cy} 
S.~Majid and H.~Ruegg, 
Phys.\ Lett.\ B {\bf 334}, 348 (1994) [arXiv:hep-th/9405107]. 

\bibitem{Lukierski:1993wx} 
J.~Lukierski, H.~Ruegg and W.~J.~Zakrzewski, 
Annals Phys.\ {\bf 243}, 90 (1995) [arXiv:hep-th/9312153]. 

\bibitem{Kowalski-Glikman:2001gp} 
J.~Kowalski-Glikman, 
Phys.\ Lett.\ A {\bf 286}, 391 (2001) [arXiv:hep-th/0102098]. 

\bibitem{Bruno:2001mw} 
N.~R.~Bruno, G.~Amelino-Camelia and J.~Kowalski-Glikman, 
Phys.\ Lett.\ B {\bf 522}, 133 (2001) [arXiv:hep-th/0107039]. 

\bibitem{Kowalski-Glikman:2002we} 
J.~Kowalski-Glikman and S.~Nowak, 
Phys.\ Lett.\ B {\bf 539}, 126 (2002) [arXiv:hep-th/0203040]. 

\bibitem{Girelli:2005dc} 
F.~Girelli, T.~Konopka, J.~Kowalski-Glikman and E.~R.~Livine, 
Phys.\ Rev.\ D {\bf 73}, 045009 (2006) [arXiv:hep-th/0512107]. 

 
 
\bibitem{Magueijo:2002xx} 
J.~Magueijo and L.~Smolin, 
Class.\ Quant.\ Grav.\ {\bf 21}, 1725 (2004) [arXiv:gr-qc/0305055]. 

\bibitem{Kimberly:2003hp} 
D.~Kimberly, J.~Magueijo and J.~Medeiros, 
Phys.\ Rev.\ D {\bf 70}, 084007 (2004) [arXiv:gr-qc/0303067]. 

\bibitem{Hinterleitner:2004ny} 
F.~Hinterleitner, 
Phys.\ Rev.\ D {\bf 71}, 025016 (2005) [arXiv:gr-qc/0409087]. 
\bibitem{Galan:2004st} 
P.~Galan and G.~A.~Mena Marugan, 
Phys.\ Rev.\ D {\bf 70}, 124003 (2004) [arXiv:gr-qc/0411089]. 

\bibitem{Amelino-Camelia:2005ne}
G.~Amelino-Camelia,
Int.\ J.\ Mod.\ Phys.\ D {\bf 14}, 2167 (2005)
[arXiv:gr-qc/0506117].

\bibitem{Galan:2005ju} 
P.~Galan and G.~A.~Mena Marugan, 
Phys.\ Rev.\ D {\bf 72}, 044019 (2005) [arXiv:gr-qc/0507098]. 


 \bibitem{Hackett:2005mb} 
J.~Hackett, 
Class.\ Quant.\ Grav.\ {\bf 23}, 3833 (2006) [arXiv:gr-qc/0509103]. 

\bibitem{Galan:2006by} 
P.~Galan and G.~A.~Mena Marugan, 
Phys.\ Rev.\ D {\bf 74}, 044035 (2006) [arXiv:gr-qc/0608061]. 

\bibitem{Ling:2006ba}
  Y.~Ling, S.~He and H.~Zhang,
  arXiv:gr-qc/0609130.

 

 
\bibitem{Hossenfelder:2005ed}
 S.~Hossenfelder, 
Class.\ Quant.\ Grav.\ {\bf 23}, 1815 (2006) [arXiv:hep-th/0510245]. 
 
 
  

 \bibitem{Kempf:1994su}
A.~Kempf, G.~Mangano and R.~B.~Mann,
Phys.\ Rev.\ D {\bf 52}, 1108 (1995).

\bibitem{ml1}
M.~Maggiore,
Phys.\ Lett.\ B {\bf 319}, 83 (1993) [arXiv:hep-th/9309034].

\bibitem{ml2}
M.~Maggiore,
Phys.\ Rev.\ D {\bf 49}, 5182 (1994) [arXiv:hep-th/9305163].

\bibitem{ml3}
A.~Camacho,
Int.\ J.\ Mod.\ Phys.\ D {\bf 12}, 1687 (2003) [arXiv:gr-qc/0305052].
 


 
\bibitem{Liberati:2004ju} 
S.~Liberati, S.~Sonego and M.~Visser, 
Phys.\ Rev.\ D {\bf 71}, 045001 (2005) [arXiv:gr-qc/0410113]. 
 
 




\bibitem{Judes:2002bw} 
S.~Judes and M.~Visser, 
Phys.\ Rev.\ D {\bf 68}, 045001 (2003) [arXiv:gr-qc/0205067]. 

\bibitem{Girelli:2004ue} 
F.~Girelli and E.~R.~Livine, 
[arXiv:gr-qc/0412004[. 

\bibitem{Girelli:2006ez} 
F.~Girelli and E.~R.~Livine, 
[arXiv:gr-qc/0612111]. 

\bibitem{Hossenfelder:2007fy} 
S.~Hossenfelder, 

\bibitem{Toller:2003tz} 
M.~Toller, 
Mod.\ Phys.\ Lett.\ A {\bf 18}, 2019 (2003) [arXiv:hep-th/0301153]. 

\bibitem{Jafari:2006rr} 
N.~Jafari and A.~Shariati, 
AIP Conf.\ Proc.\ {\bf 841}, 462 (2006) [arXiv:gr-qc/0602075]. 

\bibitem{Scargle:2006kr} 
J.~D.~Scargle, J.~P.~Norris and J.~T.~Bonnell, 
[arXiv:astro-ph/0610571]. 

\bibitem{Kosinski:2002gu} 
P.~Kosinski and P.~Maslanka, 
Phys.\ Rev.\ D {\bf 68}, 067702 (2003) [arXiv:hep-th/0211057]. 

 

\end{thebibliography}
\end{document}